\documentclass[12pt,namedreferences]{ijtp}
\usepackage{latexsym,amstext,amssymb,amsmath}

\def\slapar{\not\!\partial}\def\slaz{\not\!z}

\newcommand{\be}{\begin{equation}}
\newcommand{\ee}{\end{equation}}

\newcommand{\Cl}{\mathcal{C}\ell_D}
\newcommand{\B}{\mathcal{B}}
\newcommand{\littleint}{\mbox{\small$\int$}}
\newcommand{\littlesum}{\mbox{\small$\sum$}}

\runningtitle{Dirac Equations from Their Solutions}
\runningauthor{Kurt Just, Zbigniew Oziewicz and Erwin Sucipto}
\begin{opening}
\title{Recovery of Dirac Equations From Their
Solutions\thanks{This paper is in a final form and no version of it will
be submitted for publication elsewhere.
}}
\author{Kurt Just\thanks{The University of Arizona, Department of
Physics, Tucson AZ 85721,
just@physics.arizona.edu}}
\author{Zbigniew Oziewicz\thanks{Universidad Nacional Aut\'onoma de
M\'exico, Facultad de Estudios Superiores, C.P. 54700 Cuautitl\'an
Izcalli, Apartado Postal \# 25, Estado de M\'exico,
oziewicz@servidor.unam.mx and Uniwersytet Wroc{\l}awski, Instytut
Fizyki Teoretycznej, plac Maksa Borna 9, 50204 Wroc{\l}aw, Poland.
Supported by el Consejo Nacional de Ciencia y Tecnolog\'{\i}a (CONACyT)
de M\'exico, Grant \# 27670 E (1999-2000) and by UNAM, DGAPA, Programa
de Apoyo a Proyectos de Investigaci\'on e Innovaci\'on Tecnol\'ogica,
Proyecto IN-109599 (1999-2000). Member of Sistema Nacional de
Investigadores de M\'exico, Expediente \# 15337.}}
\author{Erwin Sucipto\thanks{University of Arizona, Department of
Physics, Tucson AZ 85721 and Institut Teknologi Bandung,
Indonesia, sucipto@physics.arizona.edu.}}\institute{}
\received{Received November 22, 1999}\end{opening}

\begin{document}\journal{ijtp}\firstpage{1}\lastpage{7}
\begin{flushright}\begin{minipage}{112mm}\rule{112mm}{0.5pt}
\noindent\footnotesize We deal with quantum field theory in the
restriction to external Bose fields. Let
$(i\gamma^\mu\partial_\mu - \B)\psi=0$ be the Dirac equation. We prove
that a non-quantized Bose field $\B$ is a functional of the Dirac field
$\psi$, whenever this $\psi$ is strictly canonical. Performing the trivial 
verification for the $\B := m = $ constant which yields the free Dirac 
field, we also prepare the tedious verifications for all 
$\B$ which are non-quantized and static. Such verifications must 
not be confused, however, with the easy and 
rigorous proof of our formula, which is shown in detail.\smallskip

\noindent\textbf{1995 Physics and Astronomy Classification Scheme:}
03.70.+k\rule{112mm}{0.5pt}\end{minipage}\end{flushright}

\hyphenation{}

\section{DIFFERENT DIRAC THEORIES} A canonical Dirac field $\psi$ is
never an
 operator, but an operator valued distribution (Jost 1965). 
For such a four component
 spinor, its hermitian conjugate
$\psi^\dagger$ and its transpose $\psi^T$, 
we postulate the anti-commutators

\begin{equation}\left[\psi(x),\psi(y)^\dagger\right]_+\delta(x^0-y^0)=
\delta(x-y)\;\;,\quad
\left[\psi(x),\psi(y)^T\right]_+\delta(x^0-y^0)=0\;\;,
\label{com}\end{equation}
where $\delta(z):=\delta(z^0)\delta(z^1)\delta(z^2)\delta(z^3)$. We
consider the Dirac equation
\begin{equation}\left\{i{\not\!\partial}^{x}-\B(x)\right\}\psi(x)
=0\quad\text{with}
\quad{\not\!\partial}^x :=\gamma^\mu\partial/\partial x^\mu
\quad \text{and}\quad\B = \gamma_0 \B^\dagger \gamma_0\label{Dirac}
\end{equation}
(the latter makes the action $\int \overline{\psi}
\left(i{\not\!\partial}-\B\right)\psi $ hermitian). Evidently, $\B$ is a
member of Dirac's Clifford algebra $\Cl$. Since this aspect is
irrelevant here, we need not choose a specific {\it representation} of
Dirac's $\gamma^\mu$ and not even a basis in $\Cl$. As another member of
this, we need the time ordered product
\begin{equation}b(x,z):= (4\pi)^2T\psi(x+z)\overline{\psi}(x-z)\quad\in
\quad \Cl\;\;.\label{b}\end{equation}

The covariant ordering needed here must directly act only on basic 
fields, not on their derivatives. Hence the time ordering of the latter 
must be {\it defined} by
\begin{equation} T\phi,_{\mu}(x)\chi(y) := 
\partial_{\mu}^{x} \;T\phi(x)\chi(y)\;\;. \nonumber \end{equation}
This (widely used, but rarely emphasized) prescription has been 
explained by (Nambu 1952), (Callan {\it et al.} 1970), (DeWitt 1984),
(Just \& The 1986), (Sterman 1993), (DeWitt - Morette 1994). 
Of the canonical relations \eqref{com}, only the first will be used here; 
but both are needed to define $\psi$ completely.

Further treatment of $\B$ and $b$ can proceed in 4 ways, of which only
the last one will be pursued here:
\begin{itemize}
\item[(a)]{} One may {\it desire} that $\B$ also be a canonical field. This
gives the usual `effective' field theory (Weinberg 1995/96). There one
starts from \eqref{com} and \eqref{Dirac} and their extensions to Bose
fields; but all these break down under
 the infinite renormalization
(Brandt 1969).
 Hence that desire, explained in the introductions of 
many books on quantum fields, is only satisfied as long as one does not 
admit interactions.
\item[(b)]{} All divergencies are prevented in Quantum Induction (QI),
where $\B$ is a non-canonical quantum field (Just \& Sucipto 1997).
For this unconventional theory, peripheral results have been explained
briefly, but only at the expense of setting aside the proofs (Just \&
Thevenot 2000, Just {\it et al.} 2000).
\item[(c)]{} Some divergencies are also avoided when one restricts $\B$
to be non-quantized forever. This is done in the mathematical theory of
heat kernels (Esposito 1998), where one studies elaborately the boundary
conditions for \eqref{b} at large separations $z$.
\item[(d)]{} In this paper, we examine a simple consequence of the
strict postulates \eqref{com} and \eqref{Dirac}. It also holds in (b),
but now we  prove it only for {\it non-quantized} $\B$ (for clarity
excluded from QI); hence the present proof holds as well for (c). We
nevertheless do not apply heat kernels, because `outer' boundary
conditions on \eqref{b} are superfluous here.
\end{itemize}
For the case (d), we prove in Section 2 the explicit recovery of $\B$
from \eqref{Dirac} as a functional of $\psi$. The result is verified for
the constant $\B = m$ in Section 3. Restricting the non-quantized $\B$
to a static $\beta \left( \vec{x}\right)$ in Section 4, we {\it prepare} its
recovery in Section 5.

\section{THE RECOVERY FORMULA}In what follows,
\begin{equation}{\slaz}^{-3}_-:=\left(z^2-i\epsilon\right)^{-2}\slaz
\qquad \text{with}\qquad\epsilon \rightarrow +0\;\;.\label{zee}
\end{equation}
For \eqref{b}, the canonical postulates \eqref{com} and \eqref{Dirac}
imply
\begin{equation}\left\{{\slapar}^x+{\slapar}^z+2i\B(x+z)\right\}b(x,z)=
2 \pi^2 \delta(z) =i{\slapar}^z{\slaz}^{-3}_-\;\;.
\label{d2}\end{equation}
Here we have used \eqref{zee} in order to give to \eqref{b} the
analyticity of a time ordered product. At this point, it will be useful
to introduce
\begin{equation}r(x,z) := b(x,z)-i{\slaz}^{-3}_- \;\;.
\label{R}\end{equation}
We shall see that this {\it remainder} is less singular than
${\slaz}^{-3}_-$ for $z\rightarrow 0$. Using \eqref{R} in \eqref{d2}, we
obtain
\begin{eqnarray}
2\B(x+z){\slaz}^{-3}_-&=&\left\{{\slapar}^x +{\slapar}^z+2i\B(x+z)\right\}
r(x,z)\label{exact}\\
& & \nonumber\\ &{\approx}&\;{\slapar}^zr(x,z)\;=\;
{\slapar}^z[b(x,z)-i{\slaz}^{-3}_-]\;\;.\label{appr}
\end{eqnarray} \vspace{2 pt}

It is essential that the {\it equalities} in \eqref{exact} and
\eqref{appr} hold strictly,
whereas the left side of \eqref{appr} only approximately equals the
right side of \eqref{exact}.
In \eqref{appr} we have used that $z\rightarrow 0$ makes the remainder
$r(x,z)$ singular,
such that the strongest singularity on the right of \eqref{exact} is
contained in ${\slapar}^zr(x,z)$. Comparing the left sides of
\eqref{exact} and \eqref{appr}, we have seen that $r(x,z)$ is less
singular than ${\slaz}^{-3}_-$; then we eliminated it by \eqref{R}. The
resulting ${\slapar}^z{\slaz}^{-3}_-= \text{const}\cdot\delta(z)$,
however, drops out when we multiply \eqref{exact} and \eqref{appr} by
$\slaz^3$, obtaining
\begin{equation}
2\B(x)+\cdots=[{\slapar}^zb(x,z)+\cdots]{\slaz}^3\;\;.
\label{exact2}
\end{equation}
The dots symbolize terms which we have neglected in \eqref{appr}
or in the approximation $\B(x+z)\approx \B(x).$ All these terms contribute
nothing to \eqref{exact2} with $z\rightarrow 0$ ; hence 
\begin{equation}
2\B(x)\;=\;\lim_{z\rightarrow 0}\;[{\slapar}^zb(x,z)]{\slaz}^3\;\;.
\label{main}
\end{equation}

While \eqref{exact2} is  a quantum field, its local limit
\eqref{main} is non-quantized, because we assumed this in \eqref{Dirac}.
Noting \eqref{b}, we see that \eqref{main} has proved
\begin{equation}
\B(x)=8\pi^2\lim_{z\rightarrow 0}\left[{\slapar}^z\;T\psi(x+z)
\overline{\psi}
(x-z)\right]{\slaz}^3= \gamma_0{\B}(x)^\dagger \gamma_0\;\;.
\label{main2}\end{equation}
The second assertion follows when we start from \eqref{d2} with the
differential operator replaced by one which acts on the bilocal field
$b(x,z)$ from the {\it right} side.

In \eqref{main2} the multiplication by ${\slaz}^3\rightarrow 0$ has
removed the singularity. Therefore, the step functions in the time
ordering need not be differentiated. Hence the ${\slapar}^z$ can be
expressed by operators acting on $x$, giving
\begin{equation}
\B(x)=8\pi^2\gamma^\mu\lim_{z\rightarrow 0}T\psi(x-z)\;
{{\stackrel{\leftrightarrow}{\partial^x_\mu}}}\;\overline{\psi}(x+z)
{\slaz}^3\;\;.
\label{main3}
\end{equation}
Here we need no longer indicate that no differentiation acts on
${\slaz}^3.$ Thus we have {\it recovered} the non-quantized Bose field with
which Dirac's equation \eqref{Dirac} has been solved, provided 
this has been done by a Dirac field $\psi$ satisfying 
\eqref{com}.\vspace{6pt}\\

In this paper we ask to what extent \eqref{main2}
can be verified by two examples:
\vspace{6 pt}
\begin{enumerate}
\item $\B=m=$constant, which yields the free $\psi$.
\item Static non-quantized $\B(x):=\beta \left(\vec{x}\right)$.
\end{enumerate}
\vspace{6 pt}
Since neither of these examples is a quantum field, the assumptions of
heat kernels are valid here (Esposito 1998). For the free Dirac field,
we verify in Section 3 the recovery of
$\B=m$ by \eqref{main2}. For non-quantized and static $\B$, 
the complete solution 
$\psi$ of \eqref{com} and \eqref{Dirac} 
follows in principle from an
eigenvalue problem in {\it three} dimensions. For such a case, we make the 
functional \eqref{main2} more specific in Section 4. The following 
Sections 3-7 describe both a very easy and an extremely difficult verification
of \eqref{main2}. Its rigorous proof (under the conditions of Section 1d)
is {\it completed} at \eqref{main}.  \\

\section{A SIMPLE VERIFICATION} 
Let us define $\delta(p,q)$ such that the {\it measure}
\begin{equation} d(p):=(2\pi)^{-3}\theta(p_0)\delta(p^2-m^2)dp
\label{measure}\end{equation}
over the sharp mass shell $ p_0=\sqrt{\vec{p}\,^2+m^2}$ makes
$$\littleint f(p)d(p)\delta(p,q) = f(q) \qquad  \text{for}\qquad
q_0=\sqrt{\vec{q}\,^2+m^2}\;\;.$$
For $m=\text{const}>0$, the non-quantized spinors $u_\sigma(p)$ with
helicity label $\sigma$ are to fulfill
\begin{equation}\nonumber
(\not\!p\mp m)\,u^{\pm}_\sigma(p) =0
\qquad\text{and}\qquad \littlesum_\sigma u_\sigma^{\pm}(p)\,
\overline{u}_\sigma^{\pm}(p)=\,\not\!p\pm m \;\;.
\end{equation}
With the Poincar\'e invariant vacuum $|\;\rangle,$ we postulate
$$ a^{\pm}_\sigma(p)|\;\rangle\,=0 \qquad \text{and}\qquad
\left[a^{\pm}_\sigma(p),{a^{\pm}_\tau(q)}^\dagger\right]_+
=\delta_{\sigma\tau}\delta(p,q)\;\;.$$
All other anti-commutators of the $a^{\pm}_\sigma(p)$ are 
assumed to vanish.

Then \eqref{com} and \eqref{Dirac}, with $\B=m=$ constant, are satisfied
by the free canonical Dirac field,
\begin{equation}
\psi(x)=\sum_\sigma\int
\left\{e^{-ipx}u^\sigma_+(p)a^+_\sigma(p)
+e^{+ipx}u^\sigma_-(p){a^-_\sigma(p)}^\dagger\right\}d(p)\;\;.
\label{fD}
\end{equation}
Its familiar propagator will be needed in the form
\begin{align}
(2\pi)^4\langle\;|T\psi(2z)\overline{\psi}(0)|\;\rangle &=i\littleint
e^{-2ipz}dp(\not\!p+i \epsilon-m)^{-1}
\qquad (\;\text{with}\; \epsilon\rightarrow +0\;)\qquad \nonumber\\
& \label{p}\\
&=\pi^2\left({i\not\!z}^{-3}_--mz^{-2}_-+\cdots\right) \;\;.\nonumber
\end{align}
Since \eqref{main2} is non-quantized, it equals its expectation value in
any state such as $|\;\rangle$. Hence \eqref{main2} is {\it verified} by
\eqref{p}, because it yields
\begin{equation}\B(x)\; = \; 8\pi^2\lim_{z\rightarrow
0}\langle\;|{\not\!\partial}^z\;T\psi(2z)\overline{\psi}
(0)|\;\rangle \;{\not\!z}^3 \; = \; m\;\;.\label{Bm} \end{equation}
No further solution of Dirac's equation \eqref{Dirac} is
known for which \eqref{main2} can be verified as easily.\\

\section{STATIC BACKGROUNDS} 
When $\B$ is not only non-quantized, but also time independent, we define
\begin{equation} 
\beta (\vec{x})\; :=\; \B(x)\;\;.
\label{s1}
\end{equation}
Let the spinors $u^\sigma(x)$ solve the eigenvalue problem
\begin{equation}
 Hu^\sigma \; =\; \omega_\sigma\cdot u^\sigma
\qquad \text{with}\qquad
H \; :=\; \gamma_0\left\{\beta(\vec{x})-i{\not\!\partial}^x\right\}\;\;.
\label{H}
\end{equation}
Although we use ${\not\!\partial}^x =\gamma^\mu\partial/\partial
x^\mu$ to avoid additional notations, \eqref{H} involves only $x^r\,\in \,
\left\{ x^1,x^2,x^3\right\}$, because $\beta$ and $u^\sigma$ are
independent of $x^0.$ Since \eqref{main2} for
\eqref{s1} makes $\beta^\dagger\gamma_0 = \gamma_0 \beta$, the operator
$H$ is {\it hermitian}. Hence its eigenvalues
$\omega_\sigma$ are real and the solutions can be made orthogonal:
\begin{equation}\littleint u^{\sigma}(\vec{x})^\dagger\,d^3x\,u^\tau
(\vec{x})=\delta^{\sigma\tau} \qquad \text{for} \qquad \sigma, \tau
\;\in\; \left\{ 1, 2, \dots\right\}\;\;.\label{ort}\end{equation}

For brevity we use notations suitable for a discrete frequency spectrum,
although that of \eqref{H} will often be continuous as in \eqref{fD}, or
mixed as in the hydrogen atom. In either case, the $\sigma$ in \eqref{H}
takes infinitely many values in contrast to \eqref{fD}, where it labels
two helicities. In addition, we assume  
\begin{equation}\littlesum_\sigma
u^\sigma(\vec{x})u^\sigma(\vec{y})^\dagger=
\delta^3(\vec{x}-\vec{y})\;\;.\label{ort2}\end{equation}
This {\it completeness} relation will be most important here. It is
compatible with \eqref{ort}, but not implied by this.
Then Dirac's equation \eqref{Dirac} is satisfied by each term of
\begin{equation}
\psi(x)=\littlesum_\sigma e^{-i\omega_\sigma
x^0}u^\sigma(\vec{x})a_\sigma\;\;.
\label{psi}
\end{equation}
We find \eqref{com} satisfied when we make \eqref{psi} a quantum
field by postulating
\begin{equation}
\left[a_\sigma,a^\dagger_\tau\right]_+=\delta_{\sigma\tau}
\qquad\text{and}\qquad
\left[a_\sigma,a_\tau\right]_+=0\;\;.
\label{a}
\end{equation}
\vspace{8 pt}
\section{DESIRABLE VERIFICATIONS} 
We specify a ground state $|\cdot\rangle$
by separating positive and negative frequencies in \eqref{psi}:
\begin{gather}
\psi(x)=\littlesum_\sigma e^{-i\omega_\sigma
x^0}u^\sigma_+(\vec{x})a_\sigma+
\littlesum_\tau e^{+i\Omega_\tau
x^0}u^\tau_-(\vec{x})b^\dagger_\tau\;\;,\nonumber\\
\label{gs}\\
a_\sigma|\cdot\rangle=0\qquad\text{and}\qquad
b_\tau|\cdot\rangle=0\;\;,\nonumber
\end{gather}
where $\omega_\sigma>0$ and $\Omega_\tau:=
-\omega_\tau>0$ . Rewriting the anti-commutators \eqref{a} in the
notation \eqref{gs}, we deduce the {\it propagator}
\begin{align}
F(x,z)&:=\;\langle\cdot|T\psi(x+z)\overline{\psi}(x-z)|\cdot\rangle
\nonumber\\ \nonumber\\
&=\;\theta(z^0)\littlesum_\sigma e^{-2i\omega_\sigma z^0}
u^\sigma_+(\vec{x}+\vec{z})\overline{u}^\sigma_+(\vec{x}-\vec{z})\nonumber\\
&\qquad\qquad 
-\theta(-z^0)\littlesum_\tau e^{+2i\Omega_\tau z^0}
u^\tau_-(\vec{x}+\vec{z})\overline{u}^\tau_-(\vec{x}-\vec{z})\;\;,
\label{p2}
\end{align}
which unlike \eqref{p} is not Poincar\'e covariant.
Having restricted the field $\beta$ in \eqref{s1} to become
non-quantized, we find it equal to its
expectation value
\begin{equation} 
\beta(\vec{x})=\,\langle\cdot|\B(x)|\cdot\rangle\,=8\pi^2
\lim_{z\rightarrow 0}
\left[{\not\!\partial}^zF(x,z)\right]{\not\!z}^3\;\;.
\label{Bs}\end{equation}
Since \eqref{main2} is the same as \eqref{main3}, we can in \eqref{Bs}
with \eqref{p2} omit those terms in which the $u^\sigma_{\pm}(\vec{x})$
are not differentiated. Returning from \eqref{gs} to the compact
notation \eqref{psi}, we obtain
\begin{multline}\beta(\vec{x})/8\pi^2=\\
\gamma^r\lim_{z\rightarrow 0}\littlesum_\sigma
u^\sigma(\vec{x}-\vec{z})
\stackrel{\leftrightarrow}{\partial^x_r}\overline{u}^\sigma(\vec{x}+\vec{z})
\{\theta(-z_0)\theta(\omega_\sigma)-\theta(z_0)\theta(-\omega_\sigma)\}
e^{2i\omega_\sigma z_0}{\not\!z}^3\;\;.
\label{Bs2}
\end{multline}

In all the limits taken in \eqref{main} through \eqref{Bs2}, $z=0$ may
be approached on any line through
Minkowski space which does not touch the cone $z^2=0.$ Hence \eqref{Bs2}
can be specialized in many ways. Starting with $\vec{z}\equiv 0,$ for
instance, we see that the matrices
$$u^\sigma(\vec{x})\stackrel{\leftrightarrow}{\partial^x_r}
\overline{u}^\sigma(\vec{x})e^{2i\omega_\sigma z_0}$$
must {\it increase} so strongly that their sums behave as $(z_0)^{-3}$ 
for $z_0\rightarrow\mp 0$ and $\omega_\sigma\rightarrow\pm\infty$.
Alternatively, we may start with $z_0\equiv\mp 0$, so that
\eqref{Bs2} simplifies to
\vspace{3 pt}
\begin{align}\beta(\vec{x})/8\pi^2&=
\gamma^r\lim_{z\rightarrow 0}\littlesum_\sigma
\theta(\omega_\sigma)u^\sigma(\vec{x}-\vec{z})
\stackrel{\leftrightarrow}{\partial^x_r}\overline{u}^\sigma(\vec{x}+\vec{z})
(\gamma_s z^s)^3\nonumber\\
& \label{Bs3}\\
&=-\gamma^r\lim_{z\rightarrow 0}\littlesum_\sigma
\theta(-\omega_\sigma)u^\sigma(\vec{x}-\vec{z})
\stackrel{\leftrightarrow}{\partial^x_r}\overline{u}^\sigma(\vec{x}+\vec{z})
(\gamma_s z^s)^3\;\;.\nonumber
\end{align}
\vspace{3 pt}\\
Here as in \eqref{gs} through \eqref{Bs2}, the sum runs either over all
solutions of \eqref{H} with frequencies $\omega_\sigma>0$ or over those
with $\omega_\sigma<0.$\\

\section{GENERAL REMARKS}
In the Coulomb field of a proton, \eqref{p2} results from all the
spinors $u^\sigma$ of either an electron or a positron. For their partly
continuous spectra, suitable {\it notations} must be invented, because
we have
for brevity used those for discrete $\omega_\sigma.$ In either case,
however, the result must verify
\begin{equation}\beta(\vec{x})=m+\gamma_0\frac{e}{|\vec{x}|}\;\;.
\label{C}\end{equation}
Since the non-quantized and static fields \eqref{s1} include the $\B = m$
of the free
Dirac field \eqref{fD}, the $\beta = m$ must also follow from
\eqref{Bs2}. However,
verifying this will be more difficult than under the manifest Lorentz
covariance employed in Section 3. The greatest obstacle to any use of
\eqref{Bs2} is that it requires {\it infinitely} many exact solutions of
\eqref{H}.

Thus we have performed one of those verifications which are possible as 
indicated in Section 5 (namely that of \eqref{C} with $e$ = 0); but we did 
so in a much simpler way. The verification shown in Section 3 consists 
of the {\it single} 
line \eqref{Bm}, because (\ref{measure} - \ref{p}) merely state our notations 
for widely familiar objects. Having tried to evaluate \eqref{Bs2} for \eqref{H} 
with $\beta(\vec{x}) = m$, we know that doing so will cost much work. Hence 
that attempt has shown that a problem which under Lorentz covariance is trivial 
can be poorly tractable when this is not manifest.

Let us also remark that all this does not concern a physical theory. 
It rather forms a didactic {\it simplification} (by non-quantized Bose fields) 
of a mathematical result from QI. This new version of Quantum Field Theory 
has only recently been suggested (Just \& Sucipto 1997). 
Hence the proof of \eqref{main3} for quantum fields $\B$ must be deferred 
until publication of QI. \\

\section{RESULTS AND EXPECTATIONS}
Whereas \eqref{C} provides one of the few simple problems in which all
solutions of \eqref{H} are known, \eqref{Bs2} must hold for every
$\beta(\vec{x})$ admitted here. For known as well as unknown $u^\sigma$, we
thus obtain the \vspace{3pt} \\

{\bf Recovery Theorem:} {\it Whenever the solutions $u^\sigma(\vec{x})$
of Dirac's equation \eqref{H} with any non-quantized and time
independent matrix
$\beta(\vec{x})\in\Cl$ fulfill the completeness relation \eqref{ort2}, that
field $\beta(\vec{x})$ is recovered by \eqref{Bs2}}.\vspace{3pt} \\

Comparing this result with the familiar `inverse scattering' theory
(Bertero \& Pike 1992),
we see that in some respect the opposite is done there. 
One wants to derive {\it approximations} to a potential by using as few as 
possible of its consequences. On the contrary, we recover $\beta(\vec{x})$ 
exactly by \eqref{Bs2}, but only when the exact solutions
$u^\sigma (\vec{x})$ of \eqref{H} are known
(either for all $\omega_\sigma >0$ or for all $\omega_\sigma < 0$). 
The further analysis of \eqref{d2} reveals that \eqref{main3} must
satisfy consistency conditions, such as Dirac induced field equations 
and the absence of Pauli terms (Just \& Thevenot 2000); 
but these do not invalidate the present results. \\

In our derivation, we have used quantum field theory (Jost 1965) in the
restriction to external Bose fields (Esposito 1998). However, the
resulting `solution' of \eqref{H} with the Bose field \eqref{s1} does
not involve quantum fields and not even time coordinates.
Thus it should equally well be of interest to readers who treat in
Dirac's equation \eqref{Dirac}
not only the matrix $\B$ but also the spinor $\psi$ as  {\it
non-quantized} fields (Thaller 1992).
For this case (in which \eqref{com} is ignored), our general result
\eqref{main3} might not be needed,
if one merely wants to derive \eqref{Bs2} from (\ref{H} - \ref{ort2}),
hence without (\ref{psi} - \ref{Bs}). Thus there remains the
\smallskip

{\bf Question:} {\it Is there a simpler way to prove \eqref{Bs2},
or will our approach remain the best method to reach that result about 
classical solutions of the time independent Dirac equation 
\eqref{H}}?

\acknowledgements For comments we are thankful to A. Borowiec, N.
Ercolani, K. Kwong, C. Levermore, W. Stoeger, and J. Thevenot.

\end{document}